\documentclass[useAMS,usenatbib]{mn2e}

\usepackage{lscape}
\usepackage{supertabular}
\usepackage{longtable}
\usepackage{graphicx}
\usepackage{txfonts}
\def\gtsima{$\; \buildrel > \over \sim \;$}
\def\ltsima{$\; \buildrel < \over \sim \;$}
\def\gsim{\lower.5ex\hbox{\gtsima}}
\def\lsim{\lower.5ex\hbox{\ltsima}}

\title[PG 1211+143]     %
  {PG 1211+143: probing high frequency lags in a high mass AGN}   %
\author[B. De Marco et al.]
  {B.~De Marco,$^{1,2}$\thanks{E-mail: demarco@iasfbo.inaf.it}
  G.~Ponti,$^3$ P.~Uttley,$^3$ M.~Cappi,$^2$
  M.~Dadina,$^2$ A. C.~Fabian,$^4$ and
  \newauthor G.~Miniutti$^5$\\ %
  $^1$Dipartimento di Astronomia, Universit\`a di Bologna, via Ranzani 1, I-40127 Bologna, Italy\\
  $^2$Istituto di Astroﬁsica Spaziale e Fisica Cosmica-Bologna, INAF, via Gobetti 101, I-40129 Bologna, Italy\\
  $^3$Faculty of Physical and Applied Science, University of Southampton, Southampton SO17 1BJ\\
  $^4$Institute of Astronomy, Madingley Road, Cambridge CB3 0HA\\
  $^5$Centro de Astrob\'iologia (CSIC-INTA), Dep. de Astrof\'isica; LAEFF, PO Box 78, E-28691, Villanueva de la Ca\~nada, Madrid, Spain}
\date{Released 2011 Xxxxx XX}

\def\LaTeX{L\kern-.36em\raise.3ex\hbox{a}\kern-.15em
    T\kern-.1667em\lower.7ex\hbox{E}\kern-.125emX}

\begin{document}


\maketitle

\begin{abstract}
We present the timing analysis of the four archived XMM-{\it Newton} observations of PG 1211+143. The source is well-known for its spectral complexity, comprising a strong soft-excess and different absorption systems. Soft energy band (0.3-0.7 keV) lags are detected over all the four observations, in the frequency range $\nu \lsim 6 \times 10^{-4}$ Hz, where hard lags, similar to those observed in black hole X-ray binaries, are usually detected in smaller mass AGN. The lag magnitude is energy-dependent, showing two distinct trends apparently connectable to the two flux levels at which the source is observed. The results are discussed in the context of disk- and/or corona-reprocessing scenarios, and of disk wind models. Similarities with the high-frequency negative lag of 1H 0707-495 are highlighted, and, if confirmed, they would support the hypothesis that the lag in PG 1211+143 represents the signature of the same underlying mechanism, whose temporal characteristics scale with the mass of the central object. 

\end{abstract}


\section{Introduction}
\label{intro}
Time lags between X-ray variations in different energy bands have been measured in a number of active galactic nuclei (AGN; e.g. Vaughan et al. 2003; McHardy et al. 2004; Ar\'{e}valo et al. 2006; Fabian et al. 2009, Zoghbi et al. 2010; Ponti et al. 2010). The lags are both Fourier-frequency and energy dependent. 

Soft-to-hard lags (i.e. hard-band variations lagging soft-band variations, conventionally having positive sign) are usually detected at relatively low frequencies (below the PSD high frequency break), their magnitudes increasing with the energy separation between the bands, with an approximately log-linear trend  (e.g. Ar\'{e}valo et al. 2006). Interestingly, the same behaviour is observed in many black hole X-ray binaries (BHXRB; e.g. Nowak et al. 1999, Miyamoto et al. 1988, 1991), thus supporting the idea that AGN are the scaled-up counter parts of smaller size black hole systems. In this scenario the main differences in timing properties would depend on the mass of the accreting object (e.g. McHardy et al. 2006; K\"{o}rding et al. 2007).

 Physical interpretations for the hard lags usually involve either Comptonization in an optically thin corona (Kazanas et al. 1997) or propagation of mass accretion rate fluctuations in the disc (Kotov et al. 2001 and Ar\'{e}valo \& Uttley 2006).

 The recent detection of a hard-to-soft high frequency time lag (i.e. soft-band variations lagging hard-band ones at high frequency, conventionally having a negative sign) in the X-ray data of 1H 0707-495, adds further complexity to the lag behaviour, revealing the presence of a distinct component, interpreted by Zoghbi et al. (2010) as a reflection reverberation signature. A different interpretation, involving distant reflection from material close to the line of sight has been proposed by Miller et al. (2010), although this conclusion has been challenged by Zoghbi et al. (2011). Another weak detection of a negative lag has been reported by Turner et al. (2011) in the 2006 XMM-{\it Newton}/\emph{Chandra} monitoring of NGC 3516, and associated to the soft energy response of the emission spectrum from a Compton-thin ionized layer of the absorbing gas. We refer also to recent results by Emmanoulopoulos et al. (2011) and Tripathi et al. (2011) on detection of soft lags in AGN.\\
\indent In this letter we present a time lag analysis of the archived XMM-{\it Newton} observations of the bright quasar PG 1211+143 ($z=0.0809$; Marziani et al. 1996). This source, while being relatively variable, has a black hole mass (M$_{BH}\sim 10^{7-8}$M$_\odot$, e.g. Kaspi et al. 2000, Peterson et al. 2004) about one order of magnitude higher than the mass of the above mentioned Seyfert galaxies for which time lags have been measured. The source is characterized by a high level of spectral complexity, and its time-averaged spectral properties have been extensively studied in the literature (e.g Pounds et al. 2003; Crummy et al. 2006; Pounds \& Reeves 2007; Reeves et al. 2008), mostly in the context of outflowing wind scenarios. Here we apply model-independent timing techniques down to the shortest possible time-scales with the intent of understanding its emission/absorption intrinsic properties.

\section{Data reduction}

XMM-{\it Newton} has observed PG 1211+143 four times, during 2001 and 2004, and during two consecutive orbits in 2007. Hereafter the two 2007 observations are referred as 2007a and 2007b, to indicate their chronological order.
All the exposures have durations in the range 51--65 ks.
Data reduction was performed using XMM Science Analysis System (SAS v. 10.0). We used data from the EPIC pn camera only, because of the highest effective area and S/N over the 0.3--10 keV energy band. 
We considered only time intervals in which the proton flares are absent, leading us to select periods with background rates $\lsim 2$cts/s. 
A circular region of 45 arcsec radius was chosen for source counts extraction, whereas two rectangular regions on the same chip of the source where used for the background. Only events with PATTERN $\leq$4 were selected. We did not find any pile-up in the data.
The spectra of each observation were grouped to a minimum of 20 counts per bin. Response matrices were generated through the RMFGEN and ARFGEN SAS tasks. Routines used for the timing analysis have been implemented using IDL v. 6.1.
\section{Light Curves and spectra}
\label{sec:lightcurves}

\begin{figure}
\centering

\includegraphics[angle=90,width=7.2cm]{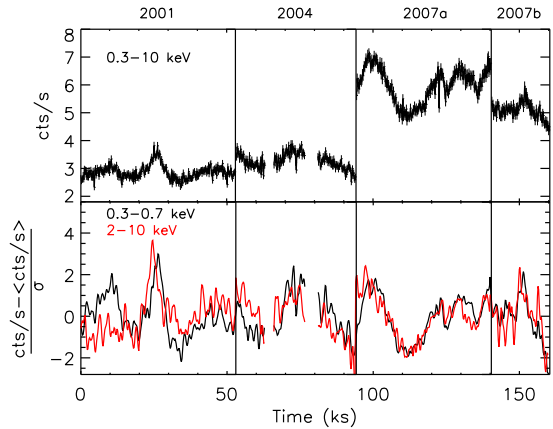}\hspace{0.5cm}

\vspace{-0.1cm}
  \caption{\emph{Upper panel}: PG 1211+143 light curves of the four XMM-{\it Newton} observations in the 0.3--10 keV energy band ($\Delta t = 200$ sec); \emph{Lower panel}: light curves in the soft (0.3-0.7 keV) and hard (2-10 keV) energy bands selected for the timing analyisis. A time shift between the two curves is observable, with the hard light curve leading the soft.}
\label{all_lc200}
    \end{figure}

The 0.3--10 keV light curves of the archived PG 1211+143 data sets are displayed in Fig. \ref{all_lc200} (upper panel), with a time binning of 200 sec. The source appears at lower fluxes during the first two observations, with the average flux increasing by a factor of $\sim$2 during the 2007a observation. The fractional root mean square (rms) variability amplitude is $\sim$7\% in the 2001 and 2004 observations and $\sim$10\% in the 2007 observation.

\begin{figure}
\centering

\includegraphics[angle=270,width=7cm]{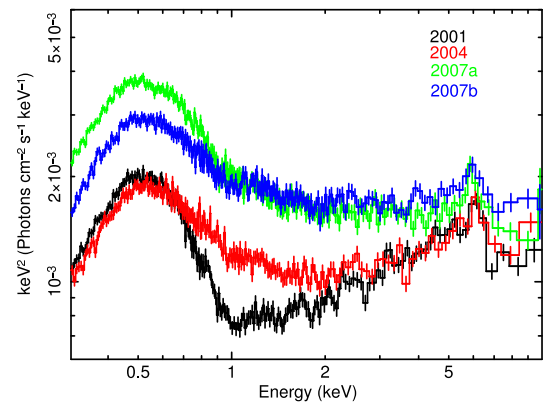}\hspace{0.5cm}

  \caption{Unfolded flux-energy spectra of all the observations. The underlying model is a constant with unit normalization.}
\label{all_specs}
    \end{figure}

 Spectra from all the observations are plotted in Fig. \ref{all_specs}. In order to visualize the relative contribution of each spectral component, they have been unfolded using a constant (with unit normalization) as the baseline model. The spectra reveal high complexity. Significant spectral variability is observed between observations.

 These data sets have been extensively studied (see Pounds et al. 2003, Kaspi \& Behar 2006, Pounds \& Reeves 2007, Reeves et al. 2008 for detailed time-averaged spectral analysis). The presence of three main spectral components is inferred: a primary power law dominating the hard X-ray spectrum, a secondary soft component producing the excess observed at energies below $\sim$1 keV, and one or more absorption systems, whose main spectral contribution falls at energies between 0.7--2 keV. The latter component is clearly visible from a comparison of the 2001 and 2004 spectra, whereby the primary continuum appears unchanged, leading Pounds \& Reeves (2007) to interprete the decrease in 0.7--2 keV flux during the 2001 observation as due to an increase in absorption covering fraction and/or column density.

 In the timing analysis described in this letter we first excluded the absorbed part of the spectrum, in order to isolate the contributions from the main hard (2--10 keV) and soft (0.3--0.7 keV) continuum components and study their temporal correlations. It is worth noting that evidence for spectral complexity at the Fe K energies has been found in PG 1211+143, mainly in the form of discrete highly ionized absorption features (e.g. Pounds et al. 2003). However, the results of our timing analysis do not significantly change if the $E>5$ keV range is neglected. The light curves in the selected soft and hard energy bands are shown in Fig.  \ref{all_lc200} (lower panel). They have been smoothed out using a gaussian kernel, so that the fluctuations due to Poisson noise are reduced. In the plot the light curves have been mean-subtracted and normalized for their standard deviation, thus allowing to directly compare the variations in the two energy bands.

\section{Timing Analysis}

\subsection{Power spectral density}
\label{PSDsec}

We estimated the power spectral density (PSD) function of PG 1211+143 by combining all the archived XMM-{\it Newton} light curves in the 0.3--10 keV energy band. Logarithmic rebinning (with 25 points/bin, Papadakis \& Lawrence 1993) and the normalization defined by Miyamoto et al. (1991) have been adopted.
The PSD is well described by a simple power law component with spectral index $\sim$1.94 (usually referred to as ``red noise''), representing the source intrinsic variability power, and dominating at $\nu \lsim 7 \times 10^{-4}$Hz, plus a constant component (also referred to as ``white noise'') resulting from the uncertainty in the measured count rate. Given the estimated mass for this object ($\sim 10^{7-8}$M$_\odot$, e.g. Kaspi et al. 2000, Peterson et al. 2004) and the steepness of its red-noise PSD, the observed range of frequencies is above the expected high frequency break, i.e. $10^{-7} \lsim \nu_B \lsim 10^{-5}$ Hz (adopting an Eddington ratio in the range Log$\ L_{Bol}/L_E\sim -0.46:0.33$, e.g. Woo \& Urry 2002). On the other hand, the power at frequencies above $\sim 7 \times 10^{-4}$Hz is dominated by the Poisson noise component.

\subsection{Lags in the frequency domain}
\begin{figure}
\centering
\vspace{1.5cm}
\begin{tabular}{p{4cm}p{4cm}p{4cm}p{4cm}}

\includegraphics[scale=0.37,angle=90]{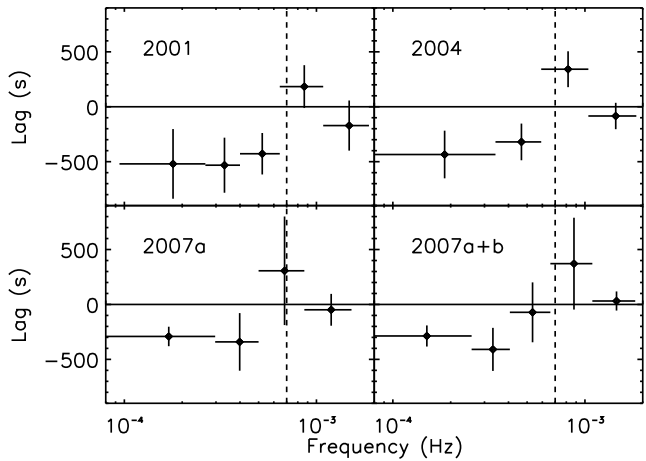} 

\vspace{-0.3cm}
\end{tabular}
  \caption{Time lag-Frequency spectra between 0.3--0.7 keV and 2--10 keV energy bands. The dashed vertical line marks the frequency threshold above which the variability is dominated by Poisson noise.}
  \label{tlag_vs_freq}

    \end{figure}

The calculation of time lags and their associated errors in the Fourier-frequency domain is carried out following Nowak et al. (1999). Results are displayed in Fig. \ref{tlag_vs_freq}.
The lags are computed between the 0.3--0.7 keV (soft excess-dominated) and the 2--10 keV (hard power law-dominated) light curves, excluding the energies where absorption dominates. In the plots, the y-axis values represent the time-shift in seconds, and it is obtained by converting the computed phase lag $\phi (f)$ into units of time through the conversion formula $\tau = \phi (f) / 2\pi f$. The frequencies are binned multiplicatively in steps of $\nu\rightarrow 1.8\nu$. We use the convention that negative lags identify a delayed response of the soft component to the hard one.

 Negative lags are observed in each of the three observations of PG 1211+143 and in the combined 2007a and 2007b data. The lags appear in the low Fourier-frequency components, i.e. $\lsim$ 6--7$ \times 10^{-4}$ Hz. This range of frequencies is below the threshold (marked with a dotted vertical line in Fig. \ref{tlag_vs_freq}) at which Poisson-noise starts to dominate the PSD (\S\ \ref{PSDsec}).

\begin{figure}
\centering

\includegraphics[scale=0.37,angle=90]{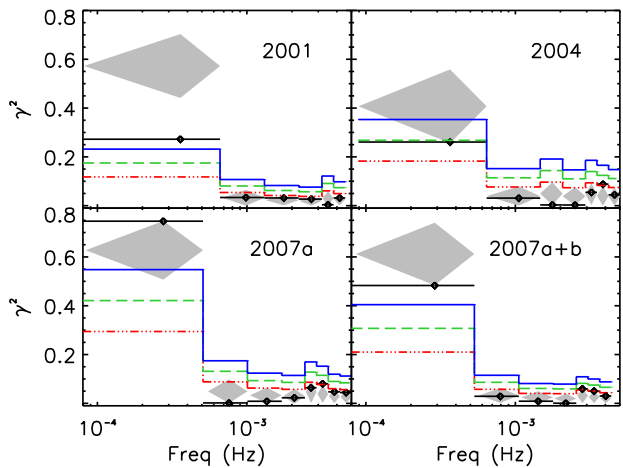}\hspace{0.5cm}

  \caption{Coherence between 0.3--0.7 keV and 2--10 keV energy bands as a function of Fourier-frequency. The $1\sigma$ (dot-dashed curve), $2\sigma$ (dashed curve), and $3\sigma$ (solid curve) contours obtained from simulations of incoherent light curves are overplotted on the data. The grey diamonds represent the 1$\sigma$ confidence intervals from simulations of coherent lightcurves.}
  \label{Coher}
    \end{figure}

\subsection{Significance of the lags}

\subsubsection{Coherence}

A necessary condition for the lags to be real is that the light curves must show some level of coherence. Coherence is a measure of the degree of linear correlation between two time series.
We computed the coherence function, using the prescription by Vaughan \& Nowak (1997). The result of coherence computation is shown in Fig. \ref{Coher}. The apparently low degree of coherence measured in PG 1211+143 light curves seems consistent with observations of other AGN, whereby the coherence has been found to decrease as the separation between the energy bands increases (e.g. Ar\'{e}valo et al. 2008). In the plots we did not apply any correction for the Poisson noise term in the cross spectrum, since the conditions given by Vaughan \& Nowak (1997) for the correction formula (equation 8) to be applicable are not satisfied, due to the fact that the measured coherence is low. Hence, we checked the coherence significance and the effects of Poisson noise using MonteCarlo simulations.

 To establish whether the observed degree of coherence is significantly different from the zero-coherence level, we simulated 1000 incoherent (i.e. uncorrelated) pairs of light curves, using the Timmer \& K\"onig (1995) algorithm, and measured their coherence functions. The incoherence of the simulated pairs is ensured by considering different realizations from the same underlying PSD (which resembles the red-noise measured one, \S\ \ref{PSDsec}, with spectral index 1.9). We added Gaussians -- with variance equal to the square of the observed count rate errors -- to each of the simulated light curves to account for the same level of Poisson noise as observed in PG 1211+143 soft and hard time series.
 For each frequency bin we measured the $1\sigma$, $2\sigma$, and $3\sigma$ spread in the distribution of simulated coherences. The resulting contour plots are shown in Fig. \ref{Coher}. The lowest frequency bin in the plots includes the frequency range where negative lags are observed. As a result the measured coherence is significantly different (at more than $3 \sigma$) from the value expected from incoherent light curves. Only during the 2004 observation does this significance decrease to $2\sigma$, probably because of the lower variability level which characterizes this observation, and the presence of gaps in the light curves, which contribute to reduce the coherence level. 

Finally, we checked whether the coherence value is intrinsically low ($\gamma^2 <1$) or whether it is decreased by the non-corrected Poisson noise contribution. In this regard we simulated 1000 coherent light curve pairs and computed the mean of the distribution of coherences -- measured including the Poisson noise term in the cross spectrum -- and its $1\sigma$ scatter. The result (see Fig. \ref{Coher}) demonstrates that only the 2001 observation has an intrinsically low degree of coherence in the lowest frequency bin, while the 2007 observations are consistent with unity-coherence. Again the observation 2004 does not allow strong conclusions to be drawn. As pointed out in \S\ \ref{sec:lightcurves}, the 2001 spectrum is characterized by the presence of a strong absorption component. This has the effect of decreasing the coherence level since the absorption system acts as a non-linear filter of the primary radiation. It is worth noting that in all the observations the coherence drops at frequencies $\nu \gsim 6 \times 10^{-4}$Hz because the uncorrelated Poisson component becomes dominant.

\subsubsection{Lag simulations}

\begin{figure}
\centering

\includegraphics[angle=90,height=3.8cm,width=7.2cm]{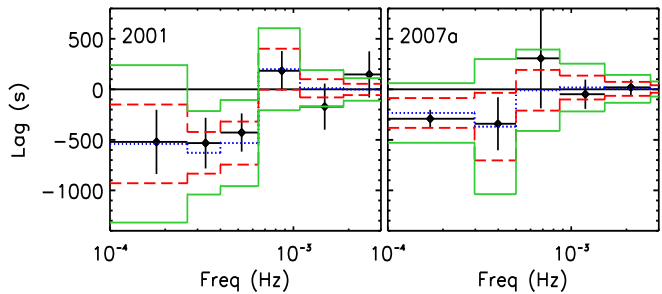}\hspace{0.5cm}

  \caption{Frequency-lag spectra of observations 2001 and 2007a. The mean value (dotted), $1\sigma$ (dashed), and $2\sigma$ (solid) contour plots as obtained from simulated light curves are overplotted on the data.}
\label{lag_cont}
    \end{figure}

The lag error bars in Fig. \ref{tlag_vs_freq} are calculated using equation (16) in Nowak et al. (1999). To check whether they are representative of the lag significance we adopted the following approach. We simulated 1000 pairs of correlated light curves with a constant time lag between them. 
The test has been done on the 2001 and 2007a data, being the longest, continuous observations, and representative of the two flux levels of the source.
A range of time lag values has been checked so as to better resemble the observed negative lags in the data. For each frequency the mean of the simulated lag distribution, and the $1\sigma$ and $2\sigma$ spreads have been computed, and displayed in Fig. \ref{lag_cont} as contour plots overplotted on real data. The contours confirm that the error bars used in the data are consistent with the $1\sigma$ dispersion in the simulated lags distribution.
 This also demonstrates that the measured lag can be well modelled by a constant lag over the entire sampled frequency range, with magnitude 650--740 sec (after correction for time dilation, $(1+z)^{-1}$, effects). The fact that the lag intrinsic magnitude is slightly higher than the observed one (as seen in the plots) is due to a red noise leak effect. Indeed, a constant time lag -- $\tau = \phi (f) / 2\pi f$ -- implies that the phase lag at each Fourier frequency is proportional to the frequency itself, with the lower frequencies showing smaller phase lags. However, this being a time series of finite length, part of the signal from the lower frequency components of the cross spectrum will leak into the higher frequency components. The result is to add a smaller phase component to the cross-spectrum, i.e. a smaller time lag, than the true value for each frequency.
 It is worth noting that the lag positive values between frequencies 0.5--1$\times 10^{-3}$ Hz are due to the phase changing its sign when it jumps from $-\pi$ to $\pi$ (see Nowak et al. 1999 for details). This is also the reason why the lag tends to zero at high frequencies. Moreover, the lag appears stable over long time-scales (of the order of years) since even the 2004 and 2007a+b lag profiles (Fig. \ref{tlag_vs_freq}) are consistent with the same constant lag magnitude and trend, further increasing the global significance of the detection.

\subsection{Lags in energy-domain}
\begin{figure}
\centering
\vspace{1.5cm}
\begin{tabular}{p{4cm}p{4cm}p{4cm}p{4cm}}

\includegraphics[scale=0.37,angle=90]{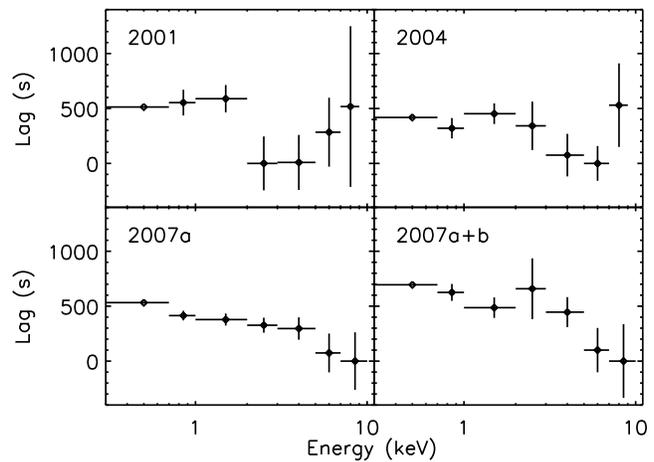} 

\vspace{-0.3cm}
\end{tabular}
  \caption{Time lag-Energy spectra over the 1--6$ \times 10^{-4}$ Hz frequency range. The lag is calculated with the energy band 0.3--0.7 keV as the reference band and then shifted so that the minimum lag corresponds to zero.}
  \label{tlag_vs_E}
    \end{figure}

To study the energy dependence of the detected time-lags over the whole energy range, $E=0.3-10$ keV, we calculated their energy spectra. The spectra are obtained by computing time lags in Fourier frequency space, between one fixed reference band and a number of adjacent energy channels, and averaging the resulting lags over the frequency interval of interest. In the present case the chosen reference band is the soft 0.3--0.7 keV band. The choice of the adjacent energy channels is arbitrary, but they need widths $\Delta E/E \sim$0.3--0.6 to statistically ensure that the variability power is above the noise level over the frequency range of interest. The integrated frequency interval includes the detected negative lags, i.e. $\Delta \nu =$ 1--6$ \times 10^{-4}$ Hz. 
The result is displayed in Fig. \ref{tlag_vs_E}, where all the lag values are shifted so that the most negative lag corresponds to zero, and the corresponding energy band becomes the reference. We note that the lag-energy spectra are similar in the 2001 and 2004 observations, while a slightly different trend is observed during the 2007 observations. The first two observations are consistent with a soft lag appearing between the middle energy band, 2--6 keV, and the soft 0.3--2 keV band. The 2007 observations, on the other hand, show a soft lag over the entire 0.3--10 keV band, characterized by an increase in magnitude towards lower energies.

\section{Discussion}

The analysis of frequency- and energy-dependent time lags in the four XMM-{\it Newton} archived observations (2001, 2004, 2007a, and 2007b observations) of the nearby quasar PG 1211+143 has been presented.
 In the timing analysis we first considered only the soft band 0.3--0.7 keV, dominated by a strong soft excess, and the hard band 2--10 keV, dominated by the power law, to avoid the range of energies strongly affected by absorption. From the measured coherence we infer that the variability registered in the two bands on Fourier-frequencies of $\sim 1-6 \times 10^{-4}$ Hz is characterized by a certain degree of linear correlation (from $\sim30\%$ to $\sim80\%$), pointing at the presence of either one unique component extending over the whole 0.3--10 keV energy range, or two distinct soft and hard correlated components. In both cases the coherent variations are diluted by incoherent variability, possibly ascribable to the absorption component. We note that the observation with stronger absorption, i.e. 2001, has lower coherence ($\gamma^2 \sim 0.3$ compared with $\gamma^2 \gsim 0.6$). The coherent variations produce the soft lag, detected in the low frequency ($\nu \lsim$ 5--6$ \times 10^{-4}$ Hz) regime during each observation. The lag is consistent with being constant over a large range of frequencies, with a magnitude of $\sim 650-740$ sec. We cannot infer whether the lag extends to higher and lower frequencies since at $\nu \gsim 6 \times 10^{-4}$ Hz the power spectrum is dominated by the white noise component, while the lower frequency part is limited by the duration and number/sampling of the observations. It is worth noting that positive hard lags are usually observed at low frequencies in Seyfert galaxies (see \S \ref{intro}). However, by rescaling their observed frequencies for the mass of PG 1211+143 they should appear below the range of frequencies probed by the current data. 

We finally checked whether the low frequency lag shows some energy dependence, finding two distinct trends appearing during the two different flux levels of the source. In the lower flux 2001 and 2004 observations the hard band leading the soft is $E\sim 2-6$ keV. On the other hand the 2007 observations are characterized by a soft lag over all the sampled energy channels. The leading energy band is at high energies and the lag magnitude increases with decreasing energy.

\subsection{An inner region origin}
The measured soft lag allows us to derive an estimate on the source and reprocessing regions geometry and distance, once the mass of the central object is known. The mass of PG 1211+143 has been estimated via reverberation mapping of the Balmer lines by Kaspi et al. (2000), $M=2.36^{+0.56}_{-0.70} \times 10^7 M_{\odot}$, and revised by Peterson et al. (2004), $M=14.6\pm 4.4 \times 10^7 M_{\odot}$.
 Hereafter the two estimates, that differ by about one order of magnitude, will be adopted as lower and upper limits to the mass of PG 1211+143.

In a reverberation scenario, the time lag we measured implies an upper limit on the primary-to-reprocessed emission regions distance of $d\lsim 2.2\times 10^{13}$cm, which corresponds to a few-to-tens $r_g$. Detection of a soft lag has been previously reported by Fabian et al. (2009), and Zoghbi et al. (2010, 2011) in the NLSy1 galaxy 1H0707-495 ($M_{BH}\sim 2 \times 10^{6} M_\odot$, Zhou \& Wang 2005). The lag was interpreted as a reverberation signature of the reflecting material in the inner regions of the accretion disc, and it was observed in the high frequency range $\Delta \nu \sim 0.8-4 \times 10^{-3}$Hz. Characteristic time-scales of an accreting system are expected to scale-up with the mass of the central object (e.g. McHardy et al. 2006; K\"{o}rding et al 2007). PG 1211+143 has an estimated black hole mass about ten-to-hundred times larger than 1H0707-495, which in turn implies that both the observed frequency of the lag and the lag magnitude should differ by a factor $\sim 10-100$. As expected, the rescaled lag frequencies ($\nu \sim 0.2-1.3 \times 10^{-4}$ Hz) and magnitude ($\tau \sim 10^2-10^3$ sec) of 1H 0707-495 yield values consistent with results found in PG 1211+143.
 These results support the idea that the same underlying physical mechanism is at work in both the sources, with the measured negative lag marking some characteristic time-scale of the system.

Nevertheless, the lag energy spectra show a puzzling behaviour which must be taken into account in discussing possible models.
 While the lower flux 2001 and 2004 spectra closely match the soft lag spectrum in 1H0707-495, and so are roughly consistent with disc reverberation (Zoghbi et al. 2011), the marginally different 2007 lag energy trend cannot be easily reconciled with this scenario.

 In this regard the time lag might be intrinsic to the source involving either a disk-corona feedback mechanism (Malzac \& Jourdain 2000), which is able to gradually soften the spectrum, or the response of a secondary, lower temperature Comptonization gas layer, producing the soft excess.

\subsection{Alternatives}
Another possibility could be that the observed lag is produced far away from the central regions, but close to the line of sight, e.g. in the absorbing gas phase observed in the X-ray spectra of PG 1211+143.
 In this case, taking the estimate given by Blustin et al. (2005) as a lower limit to the distance between the absorbing gas in PG 1211+143 and the central engine, i.e. $r\gsim 2.2-0.35 \times 10^{-3}$pc $\sim$ 300 $r_g$ (within the BLR) we infer that the gas responsible for the observed lag should be located very close ($\theta\lsim 10\deg$) to the line of sight. In this scenario the lag should be produced by the gas re-emitted/scattered component of a disc wind (e.g. Miller et al. 2010, Turner et al. 2011). As expected from numerical simulations of disk winds (e.g. Proga \& Kallman 2004, Sim et al. 2010) the ionised absorption is observed to be highly variable in PG 1211+143 (Reeves et al. 2008, Tombesi et al. 2010). Such a variable wind structure would, therefore hardly reconcile with the apparently constant lag we observe in all the observations. Moreover, the low opening angle as deduced from the lag, and the fact that the lag itself appears to be constant over a broad frequency range would imply a highly collimated structure along the line of sight which seems difficult to reproduce in wind models.

\section*{Acknowledgments}
This work is based on observations obtained with XMM-{\it Newton}, an ESA science mission with instruments and contributions directly funded by ESA Member States and NASA. GP acknowledges support via an EU Marie Curie Intra-European Fellowship under contract no. FP7-PEOPLE-2009-IEF-254279. GM thanks the Spanish Ministry of Science and Innovation for support through grants AYA2009-08059 and AYA2010-21490-C02-02. GM also thanks the Institute of Astronomy for support through its summer visitor program. The authors acknowledge the referee for helpful comments.


\end{document}